\documentclass{paper}

\usepackage{cite} 

\usepackage{graphicx}  

\usepackage{url}       

\usepackage{amsmath}   

\usepackage{amsthm}

\newcommand{\scifun}[1]{\texttt{#1}}
\newcommand{\scivar}[1]{\texttt{#1}}

\newtheorem{theorem}{Theorem}[section]
\newtheorem{proposition}[theorem]{Proposition}

\usepackage{listings}

\usepackage{url}

\lstdefinelanguage{scilabscript}%
  {morekeywords={function, endfunction, if, then, else, end, 
  real, imag, abs, number_properties, complex},%
   alsoletter=\%,
   sensitive,%
   morecomment=[l]//,%
   morestring=[b]",%
   morestring=[m]'%
   numbers=none,%
   basicstyle=\footnotesize\ttfamily ,%
   keywordstyle=\bfseries,%
   commentstyle=\bfseries,%
   showspaces=false,%
   stringstyle=\bfseries
  }[keywords,comments,strings]%

\begin{document}

\title{A Robust Complex Division in Scilab}

\author{Michael~Baudin,
        Robert~L.~Smith}

\maketitle

\begin{abstract}
The most widely used algorithm for floating point 
complex division, known as Smith's method, may fail more often than expected. 
This document presents two improved complex 
division algorithms. 
We present a proof of the robustness of the first improved 
algorithm. 
Numerical simulations show that this algorithm performs 
well in practice and is significantly more robust  
than other known implementations. 
By combining additionnal scaling methods with this first 
algorithm, we were able to create a second algorithm, 
which rarely fails.
\end{abstract}



\section{Introduction}

In floating point arithmetic, designing a robust complex division 
algorithm is surprisingly difficult. 
We mainly focus in this article on double precision IEEE 754-2008 
binary 64 floating point numbers, since these are the floating point 
numbers used in Scilab.

Indeed, it is easy to present numerical experiments where a 
naive, straightforward implementation fails to produce 
an accurate answer. 
This type of failure is caused by the 
$c^2+d^2$ term which appears in the denominator of 
the complex division expression. 
This intermediate term can overflow, which is the main 
reason why the naive method fails to work in practice for 
a large range of input values. 

Smith's 1962 method \cite{Smith1962} is based on an algorithm which 
greatly improves the robustness of complex  
division in floating point arithmetic. 
This algorithm is used in Scilab, and in many other numerical 
computing softwares as well. 
Actually, this algorithm is not as robust as we might 
think. 
It is easy to find particular complex divisions where 
Smith's method fails. 

In these cases, as we are going to see soon, the failure 
is \emph{complete} in the sense that 
none of the digits in the result are correct. 
In the current paper, we "only" try to get the correct 
magnitude in the results, but this 
is a difficult challenge in the case of the complex division. 

Many authors have recognized this fact, but we found that few authors 
actually provide better algorithms. 
Still, in 1985, Stewart \cite{214414,Stewart:1986:CNC} provided an 
algorithm which improves the robustness of Smith's algorithm, 
by considering all the ways of computing a product 
$a_1 \cdot a_2 \cdot a_3$ which appears in an intermediate 
expression of Smith's complex division. 
Stewart proved that, if $a_1,a_2$ and $a_3$ are three floating point 
numbers such that the product $a_1 \cdot a_2 \cdot a_3$ is a 
floating point number, then there is a way to compute the product which 
avoids underflows.
There are exactly three ways of computing this product, namely 
$(a_1 \cdot a_2) \cdot a_3$, $a_1 \cdot (a_2 \cdot a_3)$ and 
$a_2 \cdot (a_1 \cdot a_3)$. 
Stewart's idea was to use the appropriate expression, by 
multiplying the number of largest magnitude with the one with 
smaller magnitude. 

Stewart's algorithm works in more cases than Smith's. 
But the formulation of the algorithm is significantly more complicated. 
This increased complexity may have discouraged developers 
to use this improved algorithm. 
In practice, Smith's method is still the most widely used 
algorithm. 
Moreover, we guess that most developers did not feel the 
need for improving their algorithm, probably by the 
lack of evidence that their complex division algorithm fails 
in situations of practical importance. 
The randomized numerical experiments that we present in this paper 
may make clearer the fact that a complex division may fail more often 
than expected. 

In this article, we present an improved Smith's algorithm, 
which performs significantly better than the original 
algorithm and is significantly simpler than Stewart's algorithm. 
More precisely, our algorithm does not require to compute 
the product $a_1 \cdot a_2 \cdot a_3$ in three different 
ways. 
In fact, we prove that one of the three ways is useless, 
which simplifies the algorithm. 

In order to perform the numerical comparisons, we created 
a Scilab module which implements all the algorithms presented 
in this paper. 
This module is available at 
\begin{center}
\url{http://forge.scilab.org/index.php/p/compdiv}
\end{center}
The functions are provided under the CeCiLL, an open-source 
license which is GPL-compatible. 
The technical report \cite{BaudinSmithImpCD2011} is provided within the project, 
under the Creative Commons licence.

We compared our method with other algorithms and 
found that most algorithms which claimed for improved accuracy 
or improved performance were in fact significantly less 
accurate than expected. 
More precisely, our numerical experiments suggest that 
the rate of failure of our improved algorithm is as low 
as Stewart's, and might be 4 orders of magnitude lower than 
a naive method, and 2 orders of magnitude lower than 
the other known implementations.

This improved algorithm can still fail in a significant 
number of situations. 
We used extensive randomized numerical experiments to create 
a collection of difficult complex divisions. 
Indeed, using randomly generated complex divisions 
overcomes the problem of trying all possible combinations 
of arguments.
Indeed, this is not possible with four arguments, each being a double, 
because this would lead to a too large number of experiments. 

This work continues our analysis of the algorithms that the first 
author began in \cite{BaudinScinaive2010}. 
In this document, we presented an example of a failure of 
Smith's algorithm on an example inspired by Stewart (this example 
is presented in the section \ref{section-smithrobust}). 
Some months after the public release of this document on the web, the 
second author \cite{Smith1962} suggested to the first author 
to use a modified formula which is at the core of the algorithm 
presented in the section \ref{section-impalgo}.

This paper is organized as follows. 
In the first section, we present the naive approach and 
present particular examples of failure of this method. 
Then we present Smith's 1962 method and analyze failure 
cases of this method. 
We present the lack of reproducibility 
of some particular complex divisions, caused by the use 
of extended precision on some processors. 
We present a review of other known complex division algorithms 
and analyze their main features. 
In the second section, we present our improved algorithm 
and a proof which states that the implementation is, in 
some sense, minimum. 
We present numerical results and compare our improved implementation 
with five other algorithms. 
Then we analyze this improved algorithm and present a second, 
much more robust algorithm. 

\section{Naive and non-naive algorithms}

In this section, we present the path from a naive implementation 
of the complex division to the more robust implementation suggested 
by Smith in 1962. 

\subsection{The binary 64 doubles}
Before beginning our journey within the complex division algorithms, 
we present in this section the parameters of the floating point 
numbers which are used in Scilab, that is, we present binary 64 doubles.

In Scilab, we use double precision floating point numbers, which  
corresponds to the binary 64 floating point numbers of the IEEE-754-2008 
\cite{P754:2008:ISF} standard. 
In this floating point system, the significand makes use 
of $p=53$ binary digits of precision and 11 binary digits in the exponent. 
The parameters of the IEEE 754 double precision floating point 
system are presented in the figure~\ref{fig-doubles-properties}.

The benefits of the IEEE 754 2008 standard \cite{P754:2008:ISF} is that it 
increases the portability of programs which are making use of floating point numbers, 
and especially those written in the Scilab language. 
Indeed, on all machines where Scilab is available, the radix, the precision and 
the number of bits in the exponent are the same. 
As a consequence, the machine epsilon $\epsilon$ is always equal 
to the same value for doubles.
This is the same for the largest positive normal $\Omega$ and 
the smallest positive normal $\mu$, which are always 
equal to the values presented in the figure \ref{fig-doubles-properties}.
The situation for the smallest positive subnormal $\alpha$ is more complex, 
but, in this paper, we assume that gradual underflow is available.

\begin{figure}
\caption{Scilab IEEE 754 doubles}
\label{fig-doubles-properties}
\begin{center}
\begin{tabular}{|ll|}
\hline
Radix $\beta$ & 2  \\
Precision $p$ & 53 \\
Exponent Bits & 11 \\
$e_{min}$ & -1022 \\
$e_{max}$ & 1023 \\
Largest Normal $\Omega$ & $\left(2-\epsilon\right)\cdot 2^{1023} \approx 1.79\times 10^{308}$ \\
Smallest Normal $\mu$ & $2^{-1022} \approx 2.22\times 10^{-308}$ \\
Smallest Subnormal $\alpha$ & $2^{-1074} \approx 4.94\times 10^{-324}$ \\
Epsilon $\epsilon$ & $2^{-52} \approx 2.220\times 10^{-16}$ \\
Unit roundoff $u$ & $2^{-53} \approx 1.110\times 10^{-16}$ \\
\hline
\end{tabular}
\end{center}
\end{figure}

Doubles have intrinsic limitations which can lead to 
rounding, overflow or underflow. 
Indeed, not all mathematical numbers $x$ 
can be represented as floating point numbers. 
We denote by $fl(x)$ the floating point representation of $x$. 
In this paper, we are mostly concerned with overflow and 
underflow. 
Indeed, if $x$ is a number such that $|x|\leq \alpha/2$, 
then the floating point representation of $x$ is 
$fl(x)=0$, i.e. $x$ underflows. 
Notice that, if $\alpha/2<|x|\leq \alpha$, then 
the double which is closest to $|x|$ is $\alpha$, which implies 
that $x$ is represented by $\pm \alpha$. 
On the other hand, if $x$ is a number such that $|x|>\Omega$, 
then the floating point representation of $x$ is 
$fl(x)=Inf$, i.e. $x$ overflows.

\subsection{The naive method}

Consider the two complex numbers $a + ib$ and $c + id$, where 
$a,b,c$ and $d$ are four real numbers and $i$ is 
the imaginary number which satisfies $i^2=-1$. 
We would like to divide these two complex number, 
that is, we would like to compute $\frac{a + ib}{c + id}=e+if$, 
where $e$ and $f$ are real numbers.
We all know the equation
\begin{eqnarray}
\label{compdiv-eq-defcomplexdiv}
\frac{a + ib}{c + id} = \frac{ac + bd}{c^2 + d^2} + i \frac{bc - ad}{c^2 + d^2},
\end{eqnarray}
where, by hypothesis, $c^2 + d^2$ is nonzero.

The Scilab function \scifun{compdiv\_naive} presented in the figure \ref{fig-compdiv-algonaive} 
is a straightforward implementation of the previous formulas. 
It takes as input the complex numbers $x=a+ib$ and $y=c+id$, 
and returns $z=e+if$.

The \scifun{complex} function creates a complex number from its 
real and imaginary parts. 
In IEEE arithmetic, all invalid operations return a Nan, meaning Not-A-Number. 
In the function \scifun{compdiv\_naive}, we call the \scifun{complex} function 
to avoid to produce Nan numbers when this is possible.
In the Scilab language, the variable \scivar{\%nan} contains a quiet Nan, 
which is designed to propagate through all operations without 
signaling an exception. 
There are several simple operations which produce 
quiet Nans, including \scivar{0/0} and \scivar{\%inf-\%inf}.. 

We avoid using the expression \scivar{r=x+\%i*y} and 
use the \scifun{complex} function instead. 
Indeed, depending on the input arguments, this might produce expressions 
such as \scivar{\%inf+\%i*\%inf}.
This involves the expression \scivar{\%i*\%inf}, 
which is executed as the multiplication of 
\scivar{0+\%i} and \scivar{\%inf+\%i*0}. 
The interpreter then distributes the product, so that the real part is 
\scivar{0*\%inf-1*0}, which generates a \scivar{\%nan}.

\begin{figure}
\caption{Naive algorithm for the complex division.}
\label{fig-compdiv-algonaive}
\lstset{language=scilabscript}
\begin{lstlisting}
function z = compdiv_naive (x,y )
    a = real(x); b = imag(x);
    c = real(y); d = imag(y);
    den = c * c + d * d
    e = (a * c + b * d) / den
    f = (b * c - a * d) / den
    z = complex(e,f)
endfunction
\end{lstlisting}
\end{figure}

While the equation \ref{compdiv-eq-defcomplexdiv} is correct  
in exact arithmetic, it may fail when we consider floating point 
numbers. 
Hence, a naive implementation based on the previous 
formula can easily fail to produce an accurate result, as 
we are going to see in the next section. 

\subsection{Failure of the naive method}
\label{section-naivefailures}

Consider the complex division 
\begin{eqnarray}
\label{eq-cd-10307}
(1 + i)/(1 + i 2^{1023}) \approx 2^{-1023} - i 2^{-1023},
\end{eqnarray}
which is accurate to more than 17 decimal digits and 
is exact with doubles.
It is easy to check that the naive algorithm produces zero (for 
both the real an imaginary part) instead of the 
correct result: there is no significant digit in the result. 
Notice that both the input numbers $a=1$, $b=1$, $c=1$ and $d=2^{1023}$ and the 
outputs $e=2^{-1023}$ and $f=-2^{-1023}$ are representable as double precision
floating point numbers. 
We now focus on the floating point representation of the intermediate expressions.
The floating point representation of $den$ is $fl(den) = fl(2^{2046}) = Inf$,
because $2^{2046}$ is larger than $\Omega$. 
Then, the $e$ and $f$ variables are computed as 
\begin{eqnarray}
fl(e) 
&=& fl((1\times 1 + 1\times 2^{1023})/Inf)  = 0, \nonumber \\
fl(f) 
&=& fl((1\times 1 - 1\times 2^{1023})/Inf)  = 0. \nonumber  \\
\end{eqnarray}
Hence, the result is computed without any significant digit,
even though both the input and the output numbers are all representable as double precision
floating point numbers.

The second test involves small numbers in the denominator of the complex fraction.
Consider the complex division 
\begin{eqnarray}
\label{eq-cd-thirdtest}
(1 + i)/(2^{-1023} +  i 2^{-1023})= 2^{1023}.
\end{eqnarray}
It is easy to check that the naive algorithm produces \scivar{Inf + Nan i} instead 
of the correct result. 
To get this result, we must first use the statement \scifun{ieee(2)}, which configures 
Scilab so that Inf and Nan numbers are generated instead 
of Scilab error messages when divisions by zero are performed. 

In the previous case, we have $a=1$, $b=1$, $c=2^{-1023}$ and $d=2^{-1023}$.
This leads to
\begin{eqnarray}
fl(den) 
&=& fl(2^{-2046} + 2^{-2046})  = 0, \nonumber \\
fl(e) 
&=& fl((1\times 2^{-1023} + 1\times 2^{-1023})/0) = Inf,  \nonumber \\
fl(f) 
&=& fl((1\times 2^{-1023} - 1\times 2^{-1023})/0)  \nonumber \\
&=& fl(0/0) = Nan. \nonumber
\end{eqnarray}

The two previous complex divisions can be computed accurately if we use 
Smith's algorithm, which is presented below.


\subsection{Smith's method (1962)}

In this section, we analyze Smith's method, which produces a more accurate 
division of two complex numbers. 

\index{Smith, Robert}
In Scilab, the algorithm for the complex 
division is done by the \emph{wwdiv} routine, which implements  
Smith's method \cite{Smith1962} in the Fortran language.
Smith's algorithm is based on normalization, which calculates the complex 
division even if the input terms are large or small. 

The starting point of the method is the mathematical definition \ref{compdiv-eq-defcomplexdiv}.
We have seen that the term $c^2 + d^2$ may generate overflows or underflows.
This is caused by intermediate expressions whose magnitudes may be larger than necessary.
The previous numerical experiments suggest that, provided that we 
had simplified the calculation, the intermediate expressions 
would not have been unnecessary large. 
Hence, Smith's idea consists in rewriting the expressions 
so that we avoid writing the term $c^2 + d^2$.
More precisely, if the magnitude of $c$ is greater than the magnitude of $d$, 
then the original numerator and denominator are multiplied 
by $1 - i(d/c)$ instead of the usual $c - id$.

\begin{figure}
\caption{Smith's algorithm for the complex division.}
\label{fig-compdiv-algosmith}
\lstset{language=scilabscript}
\begin{lstlisting}
function z = compdiv_smith ( x , y )
    a = real(x); b = imag(x);
    c = real(y); d = imag(y);
    if ( abs(d) <= abs(c) ) then
        r = d/c
        den = c + d * r
        e = (a + b * r) / den
        f = (b - a * r) / den
    else
        r = c/d
        den = c * r + d
        e = (a * r + b) / den
        f = (b * r - a) / den
    end
    z = complex(e,f)
endfunction
\end{lstlisting}
\end{figure}

The \scifun{compdiv\_smith} function presented in figure \ref{fig-compdiv-algosmith} 
implements Smith's method in the Scilab language.
It is easy to check that Smith's method performs very well for the relatively 
difficult complex divisions that we met earlier. 

\subsection{Robustness of Smith's algorithm}
\label{section-smithrobust}

In this section, we present an example where Smith's method does not perform as 
expected.

The following example is inspired by an example by Stewart \cite{214414}. 
While Stewart gives an example based on a machine with an exponent range 
$\pm 99$, we consider an example which is based on Scilab's doubles. 

Consider the complex division
\begin{eqnarray}
\label{eq-smithdiffcase0}
(2^{1023} +  i 2^{-1023})/(2^{677} +   i 2^{-677})
\approx 2^{346} -  i 2^{-1008},
\end{eqnarray}
which is accurate to more than 17 decimal digits and is exact with 
doubles. 
The following Scilab script compares the naive implementation, Smith's method
and Scilab's division operator.
The session is performed with Scilab v5.2.0 under a 32 bits Windows
using a Intel Xeon processor.
\lstset{language=scilabscript}
\begin{lstlisting}
-->x = 2^1023 + %i * 2^-1023;
-->y = 2^677 + %i * 2^-677;
-->z = 2^346 - %i * 2^-1008;
-->z = compdiv_naive ( x , y )
 z  =
    Nan 
-->z = compdiv_smith ( x , y )
 z  =
    1.43D+104 
-->x/y
 ans  =
    1.43D+104 - 3.64D-304i
\end{lstlisting}
In the previous case, the naive implementation does not produce any correct digit, as 
expected. 
Smith's method, produces a correct real part, but an inaccurate imaginary 
part. 
Once again, Scilab's division operator provides the correct answer.

We check that Smith's algorithm is not accurate in this case. 
We have $a=2^{1023}$, $b=2^{-1023}$, $c=2^{677}$ and $d=2^{-677}$ and 
the algorithm performs the following steps.
\begin{eqnarray}
fl(r) 
&=& fl(2^{-677} / 2^{677}) = fl(2^{-1354}) = 0 \nonumber \\
fl(den)  
&=& fl(2^{677} + 2^{-677} \times 0)  = 2^{677} \nonumber \\
fl(e) 
&=& fl((2^{1023} + 2^{-1023} \times 0) / 2^{677}) \nonumber \\
&=& fl(2^{1023} / 2^{677})  = fl(2^{346}) \nonumber \\
&\approx& 1.43 \times 10^{104} \nonumber \\
fl(f) 
&=& fl((2^{-1023} - 2^{1023} \times 0) / 2^{677}) \nonumber  \\
&=& fl(2^{-1023} / 2^{677})  = fl(2^{-1700}) \nonumber \\
&=& 0 \nonumber 
\end{eqnarray}
We see that the variable $r$ underflows, so that it is 
represented by zero. 
This simplifies the denominator $den$,
but this variable is still correctly computed, because it is dominated 
by the term $c=2^{677}$. 
The real part $e$ is still accurate, because,
once again, the computation is dominated by the term $a$.
The imaginary part $f$ is wrong, because this term should be 
dominated by the term $ar$. 
Since $r$ underflows, it 
is represented by zero, which completely changes the result of the 
expression \scivar{b-a*r}, which is now equal to $b=2^{-1023}$.
Therefore, the result is equal to $2^{-1023} / 2^{677}$, which 
underflows to zero. 

Numerical experiments indicate that the most common cause of failure 
of Smith's algorithm is produced by the underflow of products such as 
\scivar{a*r} or \scivar{b*r}.
This is a direct consequence of the inequality $|r|\leq 1$, which implies 
that $r$ the probability that underflows is high. 
But not all underflows of $r$ are producing large relative errors. 
Assume, for example, that $|d|\leq |c|$. 
It can be proved \cite{BaudinSmithImpCD2011} that the underflow of the 
produce \scivar{d*r} in the evaluation of \scivar{den} does not produce, 
by itself, a large relative error.
On the other hand, it can be proved that the absolute error of $e$ 
and $f$ can be large if both \scivar{a*r}, \scivar{b*r} and 
\scivar{d*r} underflow. 
This problem is the motivation for an improved algorithm which 
is presented in the section \ref{section-impalgo}.

Since Scilab makes use of Smith's formula, it should fail in this 
case. 
But we have seen that Scilab is still able to perform accurately.
This unexpected accuracy is analyzed in the next section. 


\subsection{Extended precision}

In this section, we analyze the cause of the good behavior of 
Scilab's division operator in the previous case. 

The figure \ref{fig-compdiv-weird} presents the results of the 
previous complex division on several numerical computation softwares, 
including Scilab and Octave.
Here, the exact result is \scivar{Exact=} \scivar{1.433+104-3.646-304i}, 
while the approximate result is \scivar{Approx.=1.433+104}.

The cause for the differences in the results is the use of extended precision 
by some processors, depending on the compilation options and depending on the 
platform \cite{MullerEtAl2010,Monniaux2008}. 
Indeed, processors of the IA32 architecture (Intel 386, 486, Pentium and 
compatibles) have a "x87" floating point unit, which has 80-bit registers. 
This "double extended" format uses a 64-bit mantissa and a 15-bit exponent. 
In general, when the compiler generates the code for the IA32 platform, 
temporary variables are stored in the x87 registers. 
The range, from $2^{-16382}\approx 10^{-4932}$ to $2^{16383}\approx 10^{4931}$, 
of the double extended format is much larger than the range 
of the binary 64 doubles and, therefore, protects the algorithms against 
potential underflows or overflows. 

Still, it is possible to perform computations based on strict doubles if 
we use the SSE2 extensions. 
These extensions introduced one 128-bit packed floating-point 
data type, which consists of two IEEE binary 64 floating-point 
numbers. 

\begin{figure}
\caption{
Result of $(2^{1023}+i\cdot 2^{-1023})/(2^{677}+i\cdot 2^{-677})$.
}
\label{fig-compdiv-weird}
\begin{center}
\begin{tabular}{l|l|l}
\textbf{Software} & \textbf{Operating System} & \textbf{Result} \\
\hline
Scilab v5.2.0 Release & Windows 32 bits & \scivar{Exact} \\
Scilab v5.2.0 Debug   & Windows 32 bits & \scivar{Approx.} \\
Scilab v5.2.0 Release & Windows 64 bits & \scivar{Approx.} \\
Scilab v5.2.0 Release & Linux 32 bits   & \scivar{Exact} \\
Octave v3.0.3         & Windows 32 bits & \scivar{Approx.} \\
Octave v3.2.4         & Windows 32 bits & \scivar{Exact}
\end{tabular}
\end{center}
\end{figure}

Depending on the compilers options used to generate the binary,
the computer may use either the SSE unit (with binary 64 floats), or the 
x87 unit (with 80-bits registers). 
On Linux, Scilab is compiled with the Gnu compiler, gcc \cite{GCCManual2008}.  
The default value of the -mfpmath option for i386 machines 
is to use the x87 floating point co-processor. 
On the other hand, on x86\_64 machines, the default uses the SSE instruction set. 
On Windows, Scilab 5.2.0 is compiled with Visual Studio and the "/arch:IA32" option. 
This makes Scilab run on older Pentium computers that do not support 
SSE2 \cite{CordenKreitzerIntel2009}: here, Scilab may use the x87 unit. 
On the other hand, Scilab 5.2.0 uses the SSE2 unit on Windows 64 bits systems. 
Scilab is not compiled with the "/arch:IA32" option on Windows since 
2010 \cite{CornetARCHIA322010}.

The issue here is that not all systems provide extended precision, which is an 
optional part of the IEEE-754 standard. 
Moreover, the programmer may have troubles with programs which 
produce a different result in Debug or in Release (i.e. optimized) mode. 
Hence, it is highly desirable to use an algorithm which provides accurate results 
by using only standard doubles. 
This is why several authors suggested more robust complex division algorithms, as 
we are going to see in the next section.


\subsection{A review of algorithms}

In this section, we present an overview of four different algorithms 
for complex division. 
This analysis reveals, among other things, that many algorithms 
are using pre and post scalings to avoid unnecessary overflows or underflows. 

The limits of Smith's method have been analyzed by Stewart in \cite{214414}.
The paper separates the relative error of the complex numbers and the relative
error made on real and imaginary parts. 
Stewart's algorithm is based on a theorem which states that if $x_1,x_2, \ldots, x_n$
are $n$ floating point representable numbers, and if their product is also 
a representable floating point number, then the product 
$\min_{i=1,n}(x_i) \cdot \max_{i=1,n}(x_i)$
is also representable. 
The algorithm uses that theorem to perform a correct computation. 

Moreover, Stewart's algorithm uses the fact that $(b+ia)/(d+ic)=e-if$. 
In Stewart's algorithm, the case $|d|\leq |c|$ is managed explicitely. 
Instead, in the case $|d|>|c|$, Stewart first switches $a$ and $b$, then $c$ and $d$, 
and uses the same algorithm as for the first branch. 
In the case where a switch was performed, the statement \scivar{f=-f} is used 
to get the correct result. 
Since the switch operation and the change of the sign in \scivar{f} 
do not produce a rounding error, we can use them safely. 

First, we notice that Stewart's algorithm performs well in the 
complex division \ref{eq-smithdiffcase0}. 
This can be expected from the algorithm, which is able to prevent 
most overflows caused by multiplications. 
Stewart's algorithm is much better than Smith's algorithm. 
Unfortunately, as we are going to see later in this paper, 
Stewart's algorithm fails in many cases.

In the ISO/IEC 9899:TC3 C99 standard \cite{ProgrammingC}, 
section G.5.1 "Multiplicative operators", the authors present a \verb|_Cdivd| function 
in the C language which implements the complex division. 
According to Kahan \cite{KahanMarketing2000} (in the Appendix "Over/Underflow Undermines 
Complex Number Division in Java"), this code is due to Jim Thomas and Fred Tydeman. 
The algorithm is based on a pre-scaling of $c$ and $d$, such that their 
base 2 exponent is made zero. 
This scales the denominator $c^2+d^2$ and avoids most overflows or underflows. 
The post-scaling produces then the correct result. 
This scaling is based on a power of 2, which avoids rounding. 
Only in the case of an IEEE exception, the algorithm recomputes the division, 
taking into account for Nans and Infinities. 
The code does not defend against overflow and underflow in the calculation of the numerator.
We will show later in this paper that the C99 algorithm performs 
nearly as well as Smith's algorithm, but fails in many cases.

In \cite{567808}, Li et al. present a complex division algorithm with scaling. 
The algorithm is made of two stages, where the external stage 
scales the numerator and denominator if they are two small 
or too large, and the internal stage is Smith's algorithm. 
An error bound is presented for this algorithm, which is presented 
in the appendix B, "Smith's Complex Division Algorithm with Scaling" of the 
technical report \cite{Li2000} (but not in the paper \cite{567808}).
The scalings are designed to avoid overflows and underflows by scaling 
up or down the arguments. 
Since the scaling is based on $16=2^4$, the associated 
multiplications do not produce rounding. 
Unfortunately, Li et al.'s algorithm is unable to fix the failure that 
we have analyzed for Smith's method. 
Moreover, the scaling used in Li et al.'s algorithm  
fails in many cases, as we are going to see later in 
this paper. 

Priest published in 2004 a complex 
division algorithm based on scaling \cite{1039814}. 
This scaling is designed to avoid overflow and harmful underflow. 
The scaling requires only four floating point multiplications and 
a small amount of integer arithmetic to compute the scale 
factor. 
The scaling is based only on the values of $c$ and $d$. 
Priest shows that choosing a scaling factor close to $|c+id|^{-3/4}$ works 
well in most situations. 
As we are going to see later in this paper, Priest's algorithm 
does not perform well in many cases. 


\section{An improved algorithm}
\label{section-impalgo}

In this section, we present an improved complex division 
algorithm and prove that the algorithm is, in some sense, 
minimum. 
We first present the algorithm in the Scilab language, then 
prove that this algorithm uses the minimum number 
of branches required to evaluate a specific intermediate expression 
accurately. 
Then we present numerical experiments which indicate that this 
algorithm indeed performs significantly better than other 
known algorithms in practical situations.

\subsection{Analysis of the algorithm}

The Scilab function \scivar{compdiv\_improved} presented in the figure \ref{fig-compdiv-algoimproved} 
is an improved complex division algorithm. 

\begin{figure}
\caption{Improved algorithm for the complex division.}
\label{fig-compdiv-algoimproved}
\lstset{language=scilabscript}
\begin{lstlisting}
function z = compdiv_improved ( x, y )
    a = real(x);    b = imag(x)
    c = real(y);    d = imag(y)
    if ( abs(d) <= abs(c) ) then
        [e,f] = improved_internal(a,b,c,d)
    else
        [e,f] = improved_internal(b,a,d,c)
        f = -f
    end
    z = complex(e,f)
endfunction
function [e,f] = improved_internal(a,b,c,d)
    r = d/c
    t = 1/(c + d * r)
    if (r <> 0) then
        e = (a + b * r) * t
        f = (b - a * r) * t
    else
        e = (a + d * (b/c)) * t
        f = (b - d * (a/c)) * t
    end
endfunction
\end{lstlisting}
\end{figure}

Assume that $|d|\leq |c|$. 
In this case, Smith's method was to compute 
\begin{eqnarray}
\label{eq-way-smith0}
    e = (a + b * r) / den, \quad
    f = (b - a * r) / den,
\end{eqnarray}
where $den=c+d*r$.

We notice that we divide two times by $den$. 
Instead, we may use the formula 
\begin{eqnarray}
\label{eq-way-smith}
    e = (a + b * r) * t, \quad
    f = (b - a * r) * t,
\end{eqnarray}
where $t=1/(c+d*r)$.
This replaces two divisions by one division and two multiplications, 
which may be faster. 
On the other hand, let us consider a system with gradual underflow. 
If $|den|$ is below the underflow threshold, that 
is if $|c+d*r|\in[\alpha,\mu[$, then the 
two calculations are not the same: $t$ overflows, 
while the expression $(a+b*r)/den$ may produce a correct result. 
As we are going to see in the randomized numerical experiments of the section 
\ref{section-randomsamplings}, this does not have a significant effect 
on the overall robustness of the algorithm. 
This comes from the fact that this type of error occurs 
only with \emph{extreme} inputs, near the underflow threshold.  
Therefore, we favor the speed, in this case.

We have seen that Smith's formula may fail when 
the expression $r = d/c$ underflows. 
Then, the product $a*r$ is evaluated as zero, which may 
produce a very inaccurate value of $f= (b - a * r) * t$.
One possible approach is to consider the three different ways of 
evaluating the expression $a*r = a * (d/c)$, namely 
$a*r=a*(d/c)$, $d*(a/c)$ and $(d*a)/c$. 
Using the appropriate expression is essentially the idea of 
Stewart. 

All in all, this leads to three possible ways of evaluating $e$ and $f$: from 
the original equation 
\begin{eqnarray}
\label{eq-way1}
            e &=& (a + b * (d/c)) * t, \quad
            f = (b - a * (d/c)) * t,
\end{eqnarray}
or from 
\begin{eqnarray}
\label{eq-way2}
            e &=& (a + d * (b/c)) * t, \quad
            f = (b - d * (a/c)) * t,
\end{eqnarray}
or from
\begin{eqnarray}
\label{eq-way3}
            e &=& (a + (d * b)/c) * t, \quad
            f = (b - (d * a)/c)) * t.
\end{eqnarray}

We are going to prove that, if both the equations \ref{eq-way1} 
and \ref{eq-way2} fail to evaluate $e$ and $f$, then 
the equations \ref{eq-way3} cannot succeed. 
Hence, we are going to prove that there is no point in trying 
to evaluate the expressions in \ref{eq-way3}, as 
Stewart's algorithm does.

More precisely, the following proposition proves that there is no point 
in evaluating $e$ by the left equation in \ref{eq-way3}.

\begin{proposition}
(\emph{Improved Complex Division})
\label{prop-impcompdiv}
Assume that $\alpha>0$ is a real positive number, $\Omega>\alpha$ is a real positive number, 
$u$ is the unit roundoff and $\epsilon=2u$ is a small positive number such 
that $0<4\epsilon<1-u$ representing the machine epsilon. 
Assume that 
\begin{eqnarray}
\label{eq-assumption-epsOU}
\alpha = \frac{4\epsilon}{(1-u)\Omega}.
\end{eqnarray}
Assume that $b,c,d$ are four real numbers in the range $[\alpha,\Omega]$. 
We assume that 
\begin{eqnarray}
\label{eq-assumption-0}
            d \leq c.
\end{eqnarray}
If 
\begin{eqnarray}
\label{eq-condition-1}
            \frac{d}{c} < \alpha, \\
\label{eq-condition-2}
            \frac{b}{c} < \alpha, 
\end{eqnarray}
therefore
\begin{eqnarray}
\label{eq-condition-3d}
            \frac{bd}{c} < \alpha.
\end{eqnarray}
\end{proposition}

We emphasize that the equation \ref{eq-assumption-epsOU} is 
satisfied by the binary 64 floating point system. 
Notice that a simplified but still interesting result can be obtained if 
we ignore gradual underflow, and consider only numbers in the 
range $[\frac{1}{\Omega},\Omega]$.
Since the proof is not much more complicated 
with \ref{eq-assumption-epsOU}, we consider it directly.

Let us analyze the assumptions of the previous proposition. 
First, the assumption \ref{eq-assumption-0} states that we are in the 
case where algorithm should use the expressions \ref{eq-way1}, 
\ref{eq-way2} or \ref{eq-way3}. 
Second, the condition \ref{eq-condition-1} states that the expression 
\ref{eq-way1} may have failed to produce an accurate 
$e$, because $r=d/c$ underflows. 
Secondly, the condition \ref{eq-condition-2} states that the expression 
\ref{eq-way2} may have failed to produce an accurate 
$e$, because $b/c$ underflows.
Finally, the condition \ref{eq-condition-3d} states that the 
product $bd/c$ underflows, so that evaluating $e$ with the equation 
\ref{eq-way3} is useless. 

\begin{proof}
The proof is by contradiction. 
Let us assume that
\begin{eqnarray}
\label{eq-condition-3}
            \alpha \leq \frac{bd}{c},
\end{eqnarray}
i.e. let's assume that $bd/c$ is a floating point number. 
By the inequality \ref{eq-condition-1}, we have 
$1 < \frac{\alpha c}{d}$. 
By the inequality \ref{eq-condition-3}, we have 
\begin{eqnarray}
\label{eq-condition-3b}
            \frac{\alpha c}{d} \leq b.
\end{eqnarray}
Therefore, $1 < b$. 
We multiply both sides by $\frac{(1-u)\Omega}{4\epsilon}$, and get
\begin{eqnarray}
\label{eq-condition-1d}
            \frac{(1-u)\Omega}{4\epsilon} < b\frac{(1-u)\Omega}{4\epsilon}. 
\end{eqnarray}
Now, the condition \ref{eq-condition-2} implies 
$\frac{b}{\alpha}< c$.
From the hypothesis \ref{eq-assumption-epsOU}, we substitute 
for $\alpha$ in the previous equation, and get
\begin{eqnarray}
\label{eq-condition-2b-2}
            b\frac{(1-u)\Omega}{4\epsilon}< c.
\end{eqnarray}
We plug the inequality \ref{eq-condition-1d} into  
\ref{eq-condition-2b-2}, and get
\begin{eqnarray}
\label{eq-condition-2c}
            \frac{(1-u)\Omega}{4\epsilon} < c.
\end{eqnarray}
By hypothesis, we have $0< 4\epsilon<1-u$. 
Therefore, $1<\frac{1-u}{4\epsilon}$, 
which implies $\Omega < \frac{(1-u)\Omega}{4\epsilon}$. 
We plug the previous inequality into \ref{eq-condition-2c}, and get
$\Omega < c$.
The previous inequality is not possible, since, by hypothesis, 
$c \in[\alpha,\Omega]$. 
Hence, the inequality \ref{eq-condition-3} is false, 
which concludes the proof.
\end{proof}

The proposition \ref{prop-impcompdiv} provides the most difficult 
part of the proof. 
But it only considers the case where $b,c$ and $d$ are positive, 
does not consider the computation of $f$ and 
does not consider the case $d>c$.
In fact, it is straightforward to derive these results. 

Let us prove that the proposition \ref{prop-impcompdiv} 
can be extended to three numbers $b,c$ and $d$ in the range $[-\Omega,-\alpha]\cup [\alpha,\Omega]$. 
Let us consider the absolute values $|b|$, $|c|$ and $|d|$. 
These three numbers are in the range $[\alpha,\Omega]$, where the proposition 
\ref{prop-impcompdiv} can be applied. 
Let us assume that $|d|\leq |c|$, that $|d/c| < \alpha$ and that 
$|b/c|< \alpha$. 
Then, the proposition \ref{prop-impcompdiv} implies that 
$|bd/c|<\alpha$, i.e. the expression $(b*d)/c$ underflows.

Let us prove that there is no point 
in evaluating $f$ by the right equation in \ref{eq-way3}.
We notice that the evaluation of $f$ in the right 
part of \ref{eq-way3} involve the product $(a*d)/c$.
It is easy to see that we can apply the proposition \ref{prop-impcompdiv} 
by replacing $b$ by $a$, which leads to the result. 

Finally, the case $d>c$ is similar, since it suffices to switch first 
$a$ and $b$, then $c$ and $d$ and to take the opposite of $f$. 

The following session shows that the improved algorithm works well in the 
same case (from the section \ref{section-smithrobust}) where the original 
Smith's algorithm failed.
\lstset{language=scilabscript}
\begin{lstlisting}
-->x = 2^1023 + %i * 2^-1023;
-->y = 2^677 + %i * 2^-677;
-->z = 2^346 - %i * 2^-1008;
-->q = compdiv_improved(x,y)
 q  =
     1.43D+104 - 3.64D-304i 
\end{lstlisting}

The previous numerical experiment indicates that the improved 
algorithm perform well in the case for which it is designed. 
We now have to explore its behavior in other situations, which 
is done in the next section. 


\subsection{Several difficult complex divisions}
\label{section-difficultcases}

In this section, we present a collection of difficult complex 
divisions, and analyze the behavior of the improved algorithm 
on these particular cases. 

\begin{figure}
\caption{A collection of difficult complex divisions.}
\label{fig-compdiv-diffcasespres}
\begin{center}
\begin{tabular}{l|lll@{}}
  & x & y & x/y \\
\hline
1 & $1+i$                   & $1+i 2^{1023}$            & $2^{-1023}(1-i)$\\
2 & $1+i$                   & $2^{-1023}+i 2^{-1023}$   & $2^{1023}$ \\
3 & $2^{1023}+i 2^{-1023}$  & $2^{677}+i 2^{-677}$      & $2^{346}-i 2^{-1008}$ \\
4 & $2^{1023}+i 2^{1023}$   & $1+i$                     & $2^{1023}$ \\
5 & $2^{1020}+i 2^{-844}$   & $2^{656}+i 2^{-780}$      & $2^{364}-i 2^{-1072}$ \\
6 & $2^{-71} + i 2^{1021}$  & $2^{1001} + i   2^{-323}$ & $2^{-1072} + i   2^{20}$ \\
7 & $2^{-347}+i 2^{-54}$    & $2^{-1037}+i 2^{-1058}$   & $z_7$ \\
8 & $2^{-1074}+i 2^{-1074}$ & $2^{-1073}+i 2^{-1074}$   & $0.6+i 0.2$ \\
9 & $2^{1015}+i 2^{-989}$   & $2^{1023}+i 2^{1023}$     & $z_9$ \\
10& $2^{-622}+i 2^{-1071}$  & $2^{-343}+i 2^{-798}$     & $z_{10}$
\end{tabular}
\end{center}
\end{figure}

The figure \ref{fig-compdiv-diffcasespres} presents some difficult complex divisions. 
The real and imaginary parts of the exact result for the cases 7, 9 and 10 are
\begin{eqnarray}
z_7 &=& 3.898125604559113300 \times 10^{289}          \nonumber \\
     && + i \cdot 8.174961907852353577 \times 10^{295},   \nonumber \\
z_9 &=&0.001953125 - i \cdot 0.001953125                      \nonumber \\
z_{10} &=& 1.02951151789360578\times 10^{-84}         \nonumber \\
        && + i \cdot 6.97145987515076231\times 10^{-220}. \nonumber 
\end{eqnarray}

The accuracy $acc$ of a computed real number compared 
with an expected real number is computed with the formula 
$acc = \textrm{floor}(-\log_2 ( re ))$
where $\log_2$ is the base-2 logarithm and 
the relative error is computed with $re = |computed-expected|$ 
$/|expected|$.
With 53 bits of precision with doubles, $acc$ is an integer  
in the range $[0,53]$, where $acc=0$ corresponds to a 
completely wrong result and $ac=53$ is the maximum 
possible accuracy. 
The accuracy of a complex computed result is the minimum of the 
accuracy of the real and imaginary parts, i.e. the worst 
accuracy is displayed. 

The figure \ref{fig-compdiv-diffcases} presents the results of several complex division algorithms 
on the difficult complex divisions that we have presented.

We now analyze these particular complex divisions in more detail.

We have already analyzed the cases 1 and 2 in the section \ref{section-naivefailures}, 
and the case 3 in the section \ref{section-smithrobust}.

The case 4 is a typical failure case for Smith's algorithm,
caused by overflow in a sum in the expression $fl(e) = fl((a + b * r) / den) 
= fl(2^{1023} + 2^{1023}*1)/2 = Inf/2=Inf$. 
The same issue makes Stewart's and the Improved algorithm fail. 
The scaled algorithms, such as Li et al.'s and Priest's, work in this case. 
In Li et al.'s algorithm, the downscaling of $a$ and $b$ by the 
expressions \scivar{a=a/16} and \scivar{b=b/16} reduces the magnitude 
of these numbers. 
Then the expression \scivar{e=(b+a*r)*t} works fine, so that Li et al.'s 
algorithm produces the correct result.

The case 5 occurs when $f$ is close to the underflow threshold. 
In Smith's algorithm, the computation is: $fl(r)=fl(d/c)=fl(2^{-780}/2^{656})=0$. 
The imaginary part is then evaluated with 
$fl(e) = fl((8.52\times 10^{-255}) / (2.99\times 10^{197})) = 0$. 
In Li et al.'s algorithm, the inputs $a$ and $b$ are downscaled, but then Smith's 
algorithm fails with the same result as previously. 
In the Improved algorithm (and in Stewart's), the underflow of $r$ is avoided 
by the use of the expression \scivar{f=(b-d*(a/c))*t}, which succeeds.

The case 6 occurs when $e$ is close to the underflow threshold. 
We notice that Smith's and the Improved algorithms both succeed. 
On the other hand, we notice that Li et al.'s algorithm fails and produces a 
zero real part. 
Since Smith's algorithm succeeds, this means that the extra-scaling 
in Li et al.'s algorithm actually results into a lower accuracy. 
Indeed, Li et al.'s algorithm down scales $a$ and $b$ with the expressions 
\scivar{a=a/16} and \scivar{b=b/16}. 
When computing the real part $e$, the algorithm uses 
$fl(e) = fl((2.647\times 10^{-23}) * (4.66\times 10^{-302})) = 0$.
When Li et al.'s algorithm finally upscales the real part with \scivar{e=e*16}, the underflow 
of $e$ has already occurred and this cannot rescue the correct result anymore. 

The cases 7 and 10 will be reviewed in the section \ref{section-needforrobust}.

\begin{figure}
\caption{Result of various complex division algorithms on particular inputs.}
\label{fig-compdiv-diffcases}
\begin{center}
\begin{tabular}{l|lllllll@{}}
\bf{Case} & \bf{Naive} & \bf{Smith} & \bf{Stew.} & \bf{Li}   & \bf{C99}  & \bf{Priest} & \bf{Impr.} \\
\hline
1  &   0  &   53  &  53  &  53  &  53  &  53  &   53   \\
2  &   0  &   53  &  53  &  53  &  53  &  53  &   53   \\  
3  &   0  &   0   &  53  &  0   &  0   &  0   &   53   \\    
4  &   0  &   0   &  0   &  53  &  0   &  53  &   0    \\    
5  &   0  &   0   &  53  &  0   &  0   &  0   &   53   \\    
6  &   0  &   53  &  53  &  0   &  53  &  53  &   53   \\    
7  &   0  &   41  &  0   &  53  &  53  &  53  &   0    \\    
8  &   0  &   0   &  0   &  53  &  0   &  52  &   0    \\    
9  &   0  &   0   &  0   &  53  &  53  &  53  &   0    \\    
10 &   0  &   5   &  5   &  5   &  5   &  53  &   5    \\    
\end{tabular}
\end{center}
\end{figure}

It is interesting to see that, in Li et al.'s algorithm, the downscaling 
makes the case 4 work, but makes the case 6 fail.


\subsection{Random samplings}
\label{section-randomsamplings}

It is in fact difficult to analyze the algorithms with a case-by-case method. 
The reason is that adding a particular treatment for a potential 
failure might generate a failure in another case. 
The downscaling of $a$ and $b$ in Li et al's algorithm is an example 
of this kind of issue. 

The Linux 32 bits machine we have for these tests is making use of the x87 registers. 
Hence, Scilab's internal complex division operator "\scivar{/}", 
which makes use of Smith's 1962 algorithm, always produces the correct result in practice.  
On the other hand, a Scilab script which implements a complex division 
based on elementary real arithmetic produces 
consistently the same results, because the values are regularly stored into 
doubles. 
In this situation, the use of x87 registers and their extended precision has 
no effect on these algorithms. 
On this Linux machine, we can then consider that the complex division 
\scivar{(a+\%i*b)/(c+\%i*d)} is the "reference", while the 
Scilab-based algorithms are the algorithms under test. 
By the way, notice that the behavior of the scripts simulates the behavior of a similar 
C (or Fortran) source code which would be compiled in such a way that it uses the 
SSE registers instead of the extended precision.

With the goal of automatically finding counter-examples, we experimented the following 
brute-force search for failures. 
It is not possible to try all possible combinations of $a, b, c$ and $d$, because 
this would lead to $(2^{64})^4=2^{256}\approx 10^{77}$ combinations, which 
is much too large. 
When the space of the input variables is 
too large, it is common to perform Monte-Carlo simulations. 
The idea of our experiment is to randomly choose a sign, 
to arbitrarily pick the unit significand 1, 
and to randomly choose an exponent in the range allowed by the 
floating point binary 64 IEEE format. 

We consider divisions with $a=s_a 2^{n_a}$, $b=s_b 2^{n_b}$, $c=s_b 2^{n_c}$ and 
$d=s_d 2^{n_d}$, where $s_a,s_b,s_c$ and $s_d$ are 
uniform random integers in the set $\{-1,+1\}$, and 
$n_a,n_b,n_c,n_d$ are uniform random integers in the 
set $\{-1074,-1073,\ldots,1023\}$.
These integers correspond to the range of exponents available 
with doubles, where subnormal numbers (in the range $\{-1074,-1073,\ldots,-1023\}$) 
are included. 

We perform $N$ complex divisions, where $N$ a large integer, e.g. $N=1 000 000$. 
We count the number of times $T$ a particular algorithm, say Smith's 1962 for example, 
produces a different result from Scilab's division using the extended precision of 
x87 registers. 
Under the assumption that Scilab/x87 result is always correct, 
then $T$ is a count of the failures of Smith's 1962 algorithm. 
The probability of a failure $p$ is then estimated by $\hat{p}=T/N$. 

When the probability is not too far away from $p=0.5$, this 
method is relatively accurate at estimating a probability of failure. 
When the probability $p$ is very close to 0, then the events are 
rare so that the Monte-Carlo simulation needs more simulations in order 
to be accurate. 
We can estimate the variance of the result with
$\hat{V}(T/N) = \hat{p}(1-\hat{p})/N$.
The accuracy of the estimated probability $\hat{p}$ can then be evaluated 
by computing the confidence interval
\begin{eqnarray}
\left(T/N-1.96\sqrt{\hat{V}(T/N)} , T/N + 1.96\sqrt{\hat{V}(T/N)}\right),
\end{eqnarray}
which contains the true probability 95\% of the times.

The figure \ref{fig-compdiv-varalgos} presents the results of 
various algorithms in this numerical experiment. 
The experiments required from $N = 10 000$ to 
$N = 100 000$ simulations, depending on the failure 
probability (smaller probabilities require more simulations for a 
given accuracy).

\begin{figure}
\caption{
Result of various complex division algorithms with 
inputs $(a,b,c,d)$ having uniformly random exponents in $\{-1074,-1073,\ldots,1023\}$. 
A lower failure probability is better. 
}
\label{fig-compdiv-varalgos}
\begin{center}
\begin{tabular}{l|l|l}
\bf{Algorithm} & \bf{Failure    } & \bf{95\% Confidence}\\
\bf{         } & \bf{Probability} & \bf{Interval} \\
\hline
Naive    & 4.90e-1 & [4.88e-1,4.93e-1] \\
Smith    & 1.29e-2 & [1.26e-2,1.33e-2] \\
Li et al.& 1.31e-2 & [1.28e-2,1.35e-2] \\
Priest   & 4.04e-2 & [3.94e-2,4.15e-2] \\
C99      & 1.43e-2 & [1.40e-2,1.47e-2] \\
Stewart  & 7.79e-4 & [7.24e-4,8.34e-4] \\
Improved & 4.07e-4 & [3.67e-4,4.47e-4] \\
\end{tabular}
\end{center}
\end{figure}

We notice that the naive algorithm fails quite often, with a failure probability close 
to 50\% in this experiment! 
The Smith algorithm reduces the failure probability down 
to approximately 1\%. 
Li et al.'s algorithm only marginally improves the robustness. 
This can be expected from the algorithm, which scales up or down 
only for extreme exponents. 
The improved algorithm improves radically on Smith, reducing the 
failure probability down to approximately 1.e-4. 

Stewart's algorithm and our improved division algorithm are 
associated with the same failure probability, as expected from 
the theory. 
The difference in the probabilities displayed in the 
table \ref{fig-compdiv-varalgos} is only an effect 
of the Monte-Carlo experiment, where the probability of failure is 
itself a random variable: the confidence intervals 
do not show any significant differences between these algorithms.

\subsection{The need for a robust algorithm}
\label{section-needforrobust}

In this section, we present failure cases for the improved algorithm. 
We present the causes of the failures, and analyze ways to get the 
correct result. 
The complex divisions we consider were presented in section \ref{section-difficultcases}. 

We first consider the case 4, where the improved algorithm produces an 
infinite number, instead of the exact finite result. 
Notice that the exact result $z = 2^{1023}\approx 8.98\times 10^{307}$ is very close to $\Omega$.
The ratio $r=d/c$ achieves its maximum value, that is, $r=1$. 
Then we compute $t=1/2$ and the expression for $e$ overflows, while 
$f$ is correctly evaluated as zero. 
The cause of this failure is the evaluation of the 
expression \scivar{a+b*r}, which overflows.
We can easily check that using the expression 
\scivar{a*t+b*r*t} instead of \scivar{(a+b*r)*t} fixes the problem. 
This is because $t$ is small, in this case, so that the sum of the 
two expressions \scivar{a*t} and \scivar{b*r*t} produces the correct 
result.

Another solution is to downscale the numerator \scivar{x}, 
as in Li et al.'s algorithm. 

It is straightforward to check that using the statement \scivar{x=x/2}, 
then calling the improved division \scivar{z=compdiv\_improved(x,y)}, 
and finally back-scaling with \scivar{z=z*2} fixes the problem.
In Li et al.'s algorithm, the downscaling is done by a factor 16 instead 
of 2, even if downscaling with 2 also works.

We then consider the case 7, where the improved algorithm 
produces infinite real and imaginary parts.
The analysis of the steps of the algorithm shows that $t$ overflows, which causes 
both $e$ and $f$ to overflow. 

The fact that $t$ overflows is explained by the fact that 
$r$ is small in this case, such that $c + d \times r \approx c = 2^{-1037}$. 
But $t\approx 1/c=2^{1037}$, which is larger than $\Omega$: $t$ overflows. 
In this case, upscaling the denominator \scivar{y}, 
as done in Li et al.'s algorithm, solves the problem. 
More precisely, we compute the scaling factor \scivar{S=2/\%eps\^{\,}2}, 
upscale the denominator with \scivar{y=y*S}, 
perform an improved complex division, and finally backscale the 
result with \scivar{z=z*S}.
\lstset{language=scilabscript}
\begin{lstlisting}
-->S = 2/%eps^2;
-->y = y * S
 y  =
    2.75D-281 + 1.31D-287i  
-->z = compdiv_improved(x,y)
 z  =
    9.61D+257 + 2.01D+264i  
-->z = z * S
 z  =
    3.89D+289 + 8.17D+295i  
\end{lstlisting}

In the case 10, the improved algorithm produces 
a correct real part, but the imaginry part is evaluated as 
7.08D-220 instead of the exact result 6.97D-220.
The cause is in the evaluation of $f$, where the formula 
fails to be accurate. 
The first step which fails is the computation of the product \scivar{a*r=0}, 
which underflows. 
Hence the expression \scivar{(a*r)*t} evaluates as zero. 
We notice that, if we evaluate the product as \scivar{(a*t)*r}, 
we get \scivar{(a*t)*r=1.10D-221}, which is accurate.
Since $r\leq 1$, the case where the product \scivar{a*r} underflows 
can be expected to be not so rare.


\subsection{A robust complex division}
In this section, we present a robust complex division 
algorithm. 
This algorithm almost never fails completely, and 
produces a result which is rarely more than 1 bit 
inaccurate. 

\begin{figure}
\caption{Robust external algorithm for the complex division.}
\label{fig-compdiv-algorobust}
\lstset{language=scilabscript}
\begin{lstlisting}
function z = compdiv_robust ( x, y )
    a = real(x); b = imag(x);
    c = real(y); d = imag(y);
    AB = max(abs([a b]))
    CD = max(abs([c d]))
    B = 2
    S = 1
    OV = number_properties("huge")
    UN = number_properties("tiny")
    Be = B/%eps^2
    if ( AB >= OV/2 ) then // Scale down a, b
        x = x/2;  S = S*2;
    end
    if ( CD >= OV/2 ) then // Scale down c, d
        y = y/2;  S = S/2;
    if ( AB <= UN*B/%eps ) then // Scale up a, b
        x = x*Be; S = S/Be;
    end
    if ( CD <= UN*B/%eps ) then // Scale up c, d
        y = y*Be; S = S*Be;
    end
    z = robust_internal(x,y)
    z = z * S
endfunction
\end{lstlisting}
\end{figure}

Our robust complex division algorithm is based on two steps:
\begin{itemize}
\item in the first step, we downscale or upscale the numerator $x$ 
and the denominator $y$, only if necessary,
\item in the second step, we perform a complex division, based 
on a robust algorithm.
\end{itemize}

The first step is similar, but not identical, to the scaling 
in Li et al. algorithm. 
We scale only with powers of 2, so that this scaling does not 
produce any roundoff error. 

The \scivar{compdiv\_robust} function presented in the figure \ref{fig-compdiv-algorobust}
implements the first step.

The downscaling of $x$ (or $y$) is done only when 
$a$ or $b$ (or $c$ or $d$) is greater than \scivar{OV/2}, where 
\scivar{OV} is the overflow threshold $\Omega$. 
This is different from Li et al.'s algorithm, where the downscaling 
is done with the factor 16, instead of 2. 
Indeed, we found no theoretical and no practical reasons to 
use the scaling factor 16. 
On the theoretical side, the factor 2 guarantees that 
expressions such as $a+b*r$, for example, do not overflow. 
This is because if $|a|,|b|\leq \Omega/2$, then $|a+br|\leq \Omega$, 
since $|r|\leq 1$. 
On the practical side, the factor 16 makes the case 6 fail for 
Li et al.'s algorithm, while the factor 2 works well. 
Moreover, the factor 2 is sufficient to pass the examples created 
by Mc Laren \cite{McLaren2009}, or similar examples. 

The upscaling of $x$ (or $y$) is similar to Li et al.'s algorithm. 
This step applies when $a$ and $b$ (or $c$ and $d$) 
are smaller than \scivar{2*UN/\%eps}, where \scivar{UN} is the 
underflow threshold $\mu$ and \scivar{\%eps} is the machine epsilon $\epsilon$ 
(\scivar{2*UN/\%eps} is approximately equal to $2 \times 10^{-292}$). 
This step is useful on floating point systems where 
the policy is store-zero instead of gradual underflow. 
Indeed, Demmel analyzed \cite{Dem84,DemmelPC2011} the intermediate expressions 
involved in Smith's algorithm. 
In some cases where the inputs are close to the underflow threshold $\mu$, 
the relative error produced by Smith's algorithm is large on a 
system with store-zero, while gradual underflow may produce the 
correct result. 
Moreover, even on a system with gradual underflow, the upscaling step 
may allow one to recover the exact result which would have otherwise been lost. 

The \scivar{robust\_internal} function presented in the figure 
\ref{fig-compdiv-algointernal} implements the complex division, based 
on a robust complex division algorithm. 
The function \scivar{internal\_compreal} computes the real 
part of the complex division $(a+ib)/(c+id)$, assuming that 
$|c|\leq |d|$. 
This is essentially the same algorithm as in \scivar{compdiv\_improved}, 
where the underflows of the expression \scivar{b*r} is managed specifically. 
The \scivar{robust\_subinternal} function computes both the 
real and imaginary parts of $(a+ib)/(c+id)$, assuming that 
$|c|\leq |d|$. 
We use the fact that the imaginary part of $(a+ib)/(c+id)$ is 
equal to the real part of $(b-ia)/(c+id)$. 

\begin{figure}
\caption{Robust internal algorithm for the complex division.}
\label{fig-compdiv-algointernal}
\lstset{language=scilabscript}
\begin{lstlisting}
function z = robust_internal(x,y)
    a = real(x);    b = imag(x)
    c = real(y);    d = imag(y)
    if ( abs(d) <= abs(c) ) then
        [e,f] = robust_subinternal(a,b,c,d)
    else
        [e,f] = robust_subinternal(b,a,d,c)
        f = -f
    end
    z = complex(e,f)
endfunction
function [e,f] = robust_subinternal(a,b,c,d)
    r = d/c
    t = 1/(c + d * r)
    e = internal_compreal(a,b,c,d,r,t)
    a = -a
    f = internal_compreal(b,a,c,d,r,t)
endfunction
function e = internal_compreal(a,b,c,d,r,t)
    if (r <> 0) then
        br = b*r
        if ( br <> 0 ) then
            e = (a + br ) * t
        else
            e = a * t + (b * t) * r
        end
    else
        e = (a + d * (b/c) ) * t
    end
endfunction
\end{lstlisting}
\end{figure}


\subsection{Numerical experiments}

In this section, we analyze the numerical results 
of the robust complex division algorithm. 

We consider the difficult complex divisions that we have presented 
in the section \ref{section-difficultcases}. 
Our numerical experiments show that the \scivar{compdiv\_robust} 
function produces the maximum possible accuracy (i.e. 53 binary digits) 
for all cases 1 to 10, with the exception of the case 8 which is accurate 
to 52 bits only.
Hence, the robust complex division \scivar{compdiv\_robust} 
almost never fails to produce the exact result in these cases. 

We performed randomized experiments, with the same 
method which has been presented in the section \ref{section-randomsamplings}. 
We considered several sets of 100 000 experiments, which all lead to the 
same following result. 

The probability that the robust complex division compute less that 52 
significant digits is lower than $10^{-6}$. 
Some complex divisions did not produce the \emph{exact} result, 
which event is associated to the 95\% probability 
interval $[2.46\times 10^{-5},1.35\times 10^{-4}]$.
We emphasize that, for these difficult complex divisions,
only the last bit is wrong for the real or imaginary 
parts: all the other bits in both the real and imaginary part 
are correct up to the 52th bit.

We have found extremely rare cases, with probability lower than $10^{-6}$, 
which are associated with a failure of the robust algorithm. 
In the following example, the real part of the exact division is 
equal to the underflow threshold $\alpha$, but the result of the 
algorithm has a zero real part.
\lstset{language=scilabscript}
\begin{lstlisting}
-->x=2^-912+%i*2^-1029;
-->y=2^-122+%i*2^46;
-->z=2^-1074+%i*-2^-958;
-->compdiv_robust(x,y)
 ans  =
  - 4.10D-289i  
\end{lstlisting}
Notice that the exact real part is $2.470328229206233817... \times 10^{-324}$, 
for which the closest double is $\alpha$, although the number is itself 
stricly lower than $\alpha$.
In this case, the scaling by a power of 2 is not used, since the 
arguments are well within the bounds. 
In the \scifun{robust\_internal} function, the algorithm detects that the 
expression \scivar{b*r} underflows. 
Therefore, it uses the expression \scivar{e=a*t+(b*t)*r} where both 
the terms \scivar{(a*t)*r} and \scivar{b*t} underflow. 
In fact, both terms are lower than $\alpha/2$: 
only their \emph{sum} is slightly greater than $\alpha/2$.


\subsection{Performance}

In this section, we analyze the performances of the algorithms that we have 
presented previously.
Since they are in the Scilab language, performance can be obtained, provided that 
we vectorize the algorithm. 
On the other hand, Scilab's complex division operator is based on a compiled 
Fortran source code. 
Therefore, in order to make the comparisons clearer, we created compiled source codes in 
the ANSI C language, based on the Scilab prototypes. 
We did this for Priest's algorithm, the Improved complex division algorithm and for 
the Robust complex division algorithm.

In order to compile these source codes, we have to pay attention to the 
compilation options that we use. 
Indeed, our goal is to see the behaviour of our algorithm when we 
use binary 64 floating point numbers only: we avoid making use of the extended precision. 
The computer for this experiment is an Intel Pentium M at 2 GHz. 
We used a Linux 32 bits operating system, Ubuntu 10.04 LTS, 
where we used the GCC compiler and the \emph{-mfpmath=sse -msse2} options. 

The algorithms are tested with a dataset with size $N=1574802$ and 
random numbers $a$, $b$, $c$ and $d$ uniform in $[0,1]$. 
This dataset size has been chosen so that performing the $N$ experiments for 
one algorithm requires more than 0.3 seconds. 
The performance is measured 10 times, after which we compute the 
average time.
We compute the MCDPS performance measure, which is the number of 
Millions (i.e. $10^6$) of Complex Division Per Seconds, 
based on the average time.
The figure \ref{fig-perfcompdivMCDPS} presents the performance of 
various complex division algorithms.

\begin{figure}
\caption{
Performances of various complex algorithms. 
}
\label{fig-perfcompdivMCDPS}
\begin{center}
\begin{tabular}{|l|l|}
\hline
\textbf{Algorithm} & \textbf{MCDPS} \\
\hline
Scilab             & 4.7 \\
Compiled Improved  & 5.0 \\
Compiled Robust    & 3.2 \\
Compiled Priest    & 4.5 \\
\hline
\end{tabular}
\end{center}
\end{figure}

The compiled Improved function is a little faster 
than Scilab. 
The performance difference is about 5\%. 
We do not have an explanation for this difference and consider that this 
is negligible in practice.
The compiled Robust is slower than Scilab.  
This can be explained because the algorithm is much more 
complex. 
The performance difference is about 33\%.
The compiled Priest algorithm is a little slower than Scilab. 
The performance difference is about 5\%.


\section{Conclusion}

In this paper, we have presented two algorithms for floating point 
complex division. 
We have designed an improved algorithm which takes into account 
for potential overflows or underflows in intermediate expressions 
involved in the complex division. 
This algorithm does not explicitly manage all the 
possible ways of performing a product of three terms, as was 
done in Stewart's algorithm. 
Hence, our algorithm is significantly simpler than 
Stewart's, but provides the same numerical accuracy.

Numerical experiments indicate that the improved algorithm 
performs much better than a naive implementation and produce 
fewer errors than Smith's 1962 algorithm. 
Its robustness is the same as Stewart's algorithms, and 
is much better than Li et al.'s, Priest's or the C99 algorithms.
This can be expected in particular from Li et al.'s algorithm, 
which improves the accuracy only for extreme inputs.

Based on this preliminary algorithm, we have designed a 
robust algorithm, which, most of the times, 
produce more than 52 significant digits in both the real 
and imaginary parts. 
While this complex division algorithm is extremely 
robust, it can fail in cases where the output is 
very close to the underflow threshold $\alpha$.


\section*{Acknowledgments}

We would like to thank Bernard Hugueney and Allan Cornet who helped 
us in the analysis of the complex division portability issues. 
We thank Bruno Pin\c{c}on and Quang Huy Tran for their comments on this 
work. 
We thank James Demmel for his explanations on the complex 
division in XBLAS. 
Micha\"el Baudin thanks the Consortium Scilab - Digit\'eo for its 
support during the preparation of this paper.

\bibliographystyle{plain}
\bibliography{improved_cdiv}

\end{document}